\documentclass[oribibl]{llncs}

\usepackage{graphicx}
\usepackage{amssymb}
\usepackage{lineno}
\DeclareGraphicsExtensions{.pdf,.jpeg,.png,.jpg,.svg}
\graphicspath{{fig/}}

\usepackage[numbers,sectionbib]{natbib} % sectionbib para dejar la bibliografia como una seccion
\begin{document}

\title{A Study on the Perception of Researchers about the Application of Agile Software Development Methods in Research}

\author{Nelson Marcelo Romero Aquino\inst{1} \\ Adolfo Gustavo Serra Seca Neto\inst{2} \and Heitor Silv\'erio Lopes\inst{3}}

\authorrunning{Romero Aquino et al.}

%%%% list of authors for the TOC (use if author list has to be modified)
\tocauthor{Nelson Marcelo Romero Aquino, Adolfo Gustavo Serra Seca Neto and Heitor Silv\'erio Lopes}

\institute
{
Federal University of Technology - Paran\'a\\
Av. Sete de Setembro, 3165 - Rebou\c{c}as CEP 80230-901\\
\inst{1} \email{nmarceloromero@gmail.com}\\
\inst{2} \email{adolfo@utfpr.edu.br}\\
\inst{3} \email{hslopes@utfpr.edu.br}
}

% \institute{Princeton University, Princeton NJ 08544, USA,\\
% \email{I.Ekeland@princeton.edu},\\ WWW home page:
% \texttt{http://users/\homedir iekeland/web/welcome.html}
% \and
% Universit\'{e} de Paris-Sud,
% Laboratoire d'Analyse Num\'{e}rique, B\^{a}timent 425,\\
% F-91405 Orsay Cedex, France}

\maketitle              % typeset the title of the contribution

\begin{abstract}
Papers on Agile Software Development methods are often focused on their applicability in commercial projects or organizations. There are no current studies that we know about addressing the application of these methods in research projects. The objective of this work is to describe the perception of researchers on the application of agile software development practices and principles for research projects. A study was conducted by constructing and applying a questionnaire to Brazilian researchers of different affiliations, formation and research areas in order to obtain information about their knowledge and openness to follow agile software development principles and practices. %Answers from researchers working at universities from Brazil were collected and evaluated.\\ %about their knowledge with respect to agile methods, if they ever applied agile activities and practices to develop software for their research projects and if they agree with the agile principles. Other data were also collected, such as the programming language used by the researchers and their willing to update their knowledge about software development.\\
%Results show that researchers are open to apply Agile Software Development methods for their projects and that they already apply at least the core concepts of some agile practices. They also agree with the agile principles and exhibit openness to updating their knowledge regarding software development. The leading agile method is Scrum, followed by eXtreme Programming. The perception of researchers on applying agile software development methods on their projects is positive. There is a consensus regarding the benefits of agile development, since the participants demonstrated openness to apply agile practices and principles.
\end{abstract}

\section{Introduction}
\label{seq:introduction}
% Context
Since the arrival of Agile Software Development (ASD) approaches, the research community has been pursued to analyze their applicability in commercial environments, projects or organizations. Some works regarding this field of study are based on comparing traditional development methods with ASD \cite{Aitken2013}, others seek to study the challenges derived from the application of agile processes in traditional organizations \cite{Boehm2005} or the suitability of using ASD methods to particular environments \cite{Conforto2014,Turk2014}. To the best of our knowledge, no research has been published to evaluate the application of ASD methods and practices in research projects within universities or the openness of the researchers to apply them to projects in which it is necessary to develop some kind of software.

% Objective of the work
The objective of this work is to describe researchers knowledge regarding ASD and their openness to follow agile software development principles and practices in research projects. For this purpose, data was collected from Brazilian researchers from different backgrounds by applying a questionnaire divided in several parts, each one regarding a concrete aspect such as the agreement with the agile principles or the knowledge about agile methods.

% Work organization
This work is organized as follows: Section \ref{seq:related_work} presents related works. Section \ref{seq:methodology} describes details about the construction of the questionnaire and its application. Section \ref{seq:results} presents the results obtained and explores their implications. Section \ref{seq:conclusion} contains the conclusion and final discussions derived from this work.

\section{Related Work}
\label{seq:related_work}
Although there are no works regarding the application of ASD in research projects in particular, several papers address their application in commercial environments. In \cite{Laanti2011}, the perception of the impact of agile methods when deployed in a very large software development environment was evaluated, mainly from the viewpoint of agile transformation. The work applied a questionnaire on a population consisting of more than 1000 respondents working at Nokia from seven different countries in Europe, North America, and Asia. Among the respondents, 90\% represented the Research and Development (R\&D) area. The work concludes that ASD received very positive feedback. In the work presented by \cite{Rodriguez2012}, a survey regarding the adoption of ASD from Finnish software practitioners was conducted, gathering answers from 408 persons representing 200 different organizations. Results show that most respondents were using ASD methods and that they are often adopted in order to increase the productivity and quality of the products and services. The study also concludes that the most common reasons preventing the adoption of ASD methods are lack of knowledge and a too traditionalist culture within an organization. The work by \citep{Diel2014} addresses the knowledge regarding ASD methods by people working at commercial environments, concretely in the Brazilian market. A qualitative questionnaire was prepared and applied to 24 Information Technology professionals distributed across 5 states of Brazil. Results of the work show that although the participants are familiar with agile principles, they adopt few agile practices.

\section{Methodology}
\label{seq:methodology}

A questionnaire containing $9$ questions was devised in order to obtain information about the knowledge and the application of agile software development methods in research projects, which we consider as any project conducted with research purposes involving software development. The size of the software that is developed by the researchers was not considered relevant for this study, since the participants develop from small applications to big systems depending on the needs of their projects. The number of questions was selected aiming to keep simple the structure of the survey so that it could remain brief and user-friendly for the participants. Details about the building process of the survey and its application are discussed in this Section.

\subsection{Building the Questionnaire}
\label{seq:methodology:building_questionnaire}

The questions of the survey were built by taking into account different aspects: knowledge about agile methods, application of agile methods, agreement with the agile principles and technical information about the researcher. The topics of each question of the questionnaire are explained next.

\begin{itemize}
\item Question 1: refers to whether the participant has insight into agile methods.
\item Question 2: concerns the knowledge of the researcher about specific agile methods.
\item Question 3: is related to which agile methods the researcher has ever applied for developing software in research projects.
\item Question 4: inquires about which agile practices the researcher has ever applied in their research projects. This question is presented without mentioning the formal names of the practices since the researcher may have applied them (even partially) without knowing that they are part of some agile method.
\item Question 5: refers to the agreement of the subject with each of the Agile Principles defined by the Agile Alliance \cite{Beck2001}. Some of the Agile Principles are related to the relationship between the developers and the clients. Thus, we did not consider them for the questionnaire since they are not applicable to research projects. 
\item Question 6: this question has the same objective as Question 5 but instead of the Agile Principles from the Agile Alliance the question focuses on the Lean Software Development (LSD) principles \citet{Poppendieck2007}.
%\item Question 7: is related to the programming languages used by the researcher to develop for research.
\item Question 7: inquires about to the openness of the subject to update his knowledge regarding software development. 
\item Questions 8 and 9: gather personal information as affiliation and line of research.
\end{itemize}

\subsection{Applying the Questionnaire}
\label{seq:methodology:applying_questionnaire}

The survey was made available on-line in Portuguese. The link of the questionnaire was sent through social media and e-mail. It was applied to $20$ anonymous subjects chosen randomly from 9 laboratories from 7 Brazilian cities and one Australian city. The participants were MSc and Phd students, research professors working at universities, undergraduate students working on their final graduation projects and undergraduate students working as interns at laboratories. The education and experience of the participants, was very diverse, which was intentionally pursued in order to gather data from a very heterogeneous population. The individuals are not related to the agile community.

%%%%%%%%%%%%%%%%%%%%%%%%%%%%%%%%%%%%%%%%%%%%%%%% Results %%%%%%%%%%%%%%%%%%%%%%%%%%%%%%%%%%%%%%%%%%%%%%%%
\section{Results}
\label{seq:results}

This Section presents an analysis of the results obtained after the application of the questionnaire briefly described in Section \ref{seq:methodology}.

\begin{figure}[!hbt]
\centering
\includegraphics[scale=0.4]{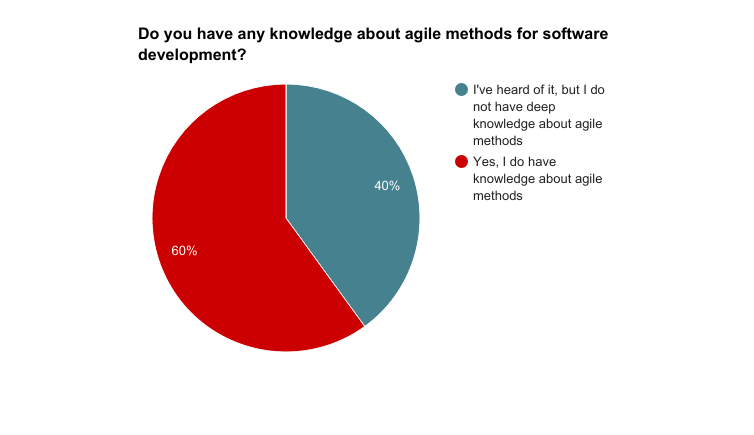}
\caption{Percentages of researchers that have knowledge about agile software development methods.}
\label{fig:0}
\end{figure}

A summary of the answers from the first question of the survey is shown in Figure \ref{fig:0}. The question refers to the basic knowledge of the participants about agile methods. Results show that 60\% of the participants have deep knowledge about agile methods whilst the other 40\% at least heard about them, although their knowledge is not too broad. There were no researchers stating that they have never heard about them, a sign of their spreading in the late years.

\begin{figure}[!hbt]
\centering
\includegraphics[scale=0.4]{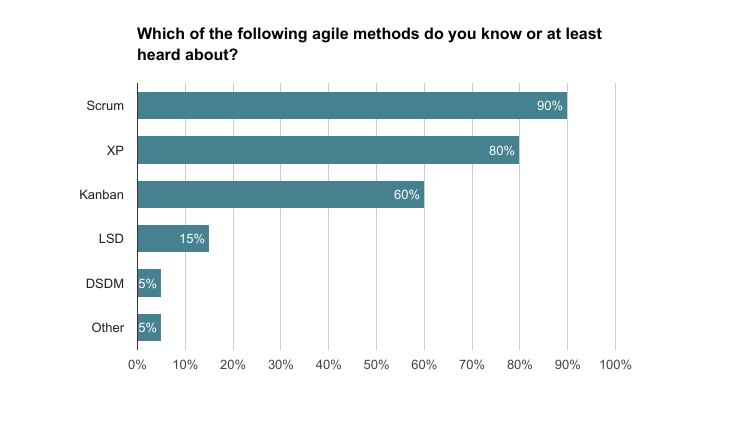}
\caption{Number of participants with knowledge about each agile method.}
\label{fig:1}
\end{figure}

Results from the second question of the questionnaire, presented in Figure \ref{fig:1}, show that the most known method is Scrum, followed by XP. The first is known by 90\% of the participants, whilst XP is known by 80\%. This is similar to the results presented in a recent survey related to the state of the agile development \cite{VersionOne2010}, which states that Scrum is currently the most used method. The study also refers to a hybrid between XP and other methods as the second most used approach, which also correspond to the result obtained in this work. Kanban is in the third place of the list, since 60\% of the participants know details or at least heard about the method. There is a significant difference between the first three methods, which we consider as the most known, and the remaining: LSD (15\%), DSDM (5\%) and Crystal (0\%). One participant selected the option \textit{Other}, adding the Planning Poker \cite{Cohn2005} practice as one of the methods he has knowledge about, although it is a practice that commonly used within the scope of XP.

\begin{figure}[!hbt]
\centering
\includegraphics[scale=0.4]{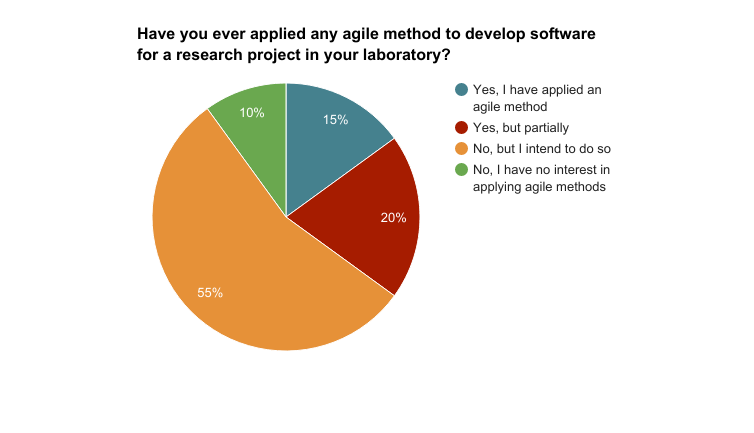}
\caption{Percentages of researchers that applied agile methods on their research projects.}
\label{fig:2}
\end{figure}

Results from the Question 3 are presented at Figure \ref{fig:2}. More than half of the participants (55\%) stated that they have never applied agile methods to develop software for their research. Notwithstanding, they intend to apply them in future projects, even though they may not have much knowledge about them. %The following 20\% of the participants stated that they have applied at least some agile software development practices, whilst the other 15\% have already applied them. The remaining 10\% never applied them and do not intend to do so.

\begin{figure}[!hbt]
\centering
\includegraphics[scale=0.4]{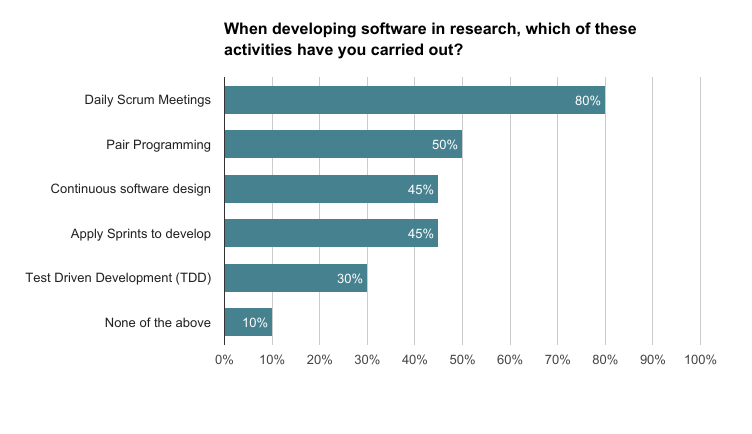}
\caption{Number of researchers that applied each agile practice in their research projects.}
\label{fig:3}
\end{figure}

For Question 4, although we present the answers in the Figure \ref{fig:3} with the formal names of the practices (e.g., Pair Programming or TDD), they were displayed in the questionnaire by using descriptions of them. Results for this question show that daily meetings to discuss issues about the software in development is a common task for the researchers. They may not be formal Daily Scrum Meetings, but the core activity (daily meetings to discuss aspects of the project) remains. Pair programming was the second most applied practice. It should also be treated as the practice discussed before: we consider that at least the core concept of the practice is applied. Incremental development (with Sprints or Iterations) and Continuous Software Design were the following practices, although with almost the same percentage as Pair Programming. Test Driven Development was the less applied practice. The use of Class Responsibility Collaborator (CRC) Cards \cite{Beck1989} for software design was also an option for this question. However, none of the participants applied it.

%Results for the questions related to the agreement with the agile principles are presented at the Figures \ref{fig:4} and \ref{fig:5}. 

\begin{figure}[!hbt]
\centering
\includegraphics[scale=0.4]{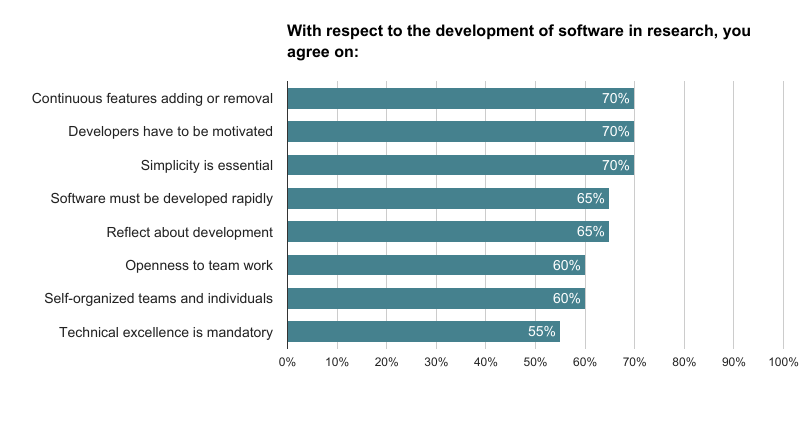}
\caption{Number of participants that agree with each agile principle of the Agile Alliance.}
\label{fig:4}
\end{figure}

Figure \ref{fig:4} presents results for the Question 5, which regards to the Agile Principles defined by the Agile Alliance \cite{Beck2001}. At least more than 50\% of the participants agree with each principle. The principle with less researchers support was the one stating that technical excellence is mandatory (55\%) whilst the most supported principles are those related to the simplicity, continuous integration and motivation for develop, all three are supported by 70\% of the participants. This shows that researchers tend to be pragmatic and that their projects requirements are usually dynamic.

\begin{figure}[!hbt]
\centering
\includegraphics[scale=0.4]{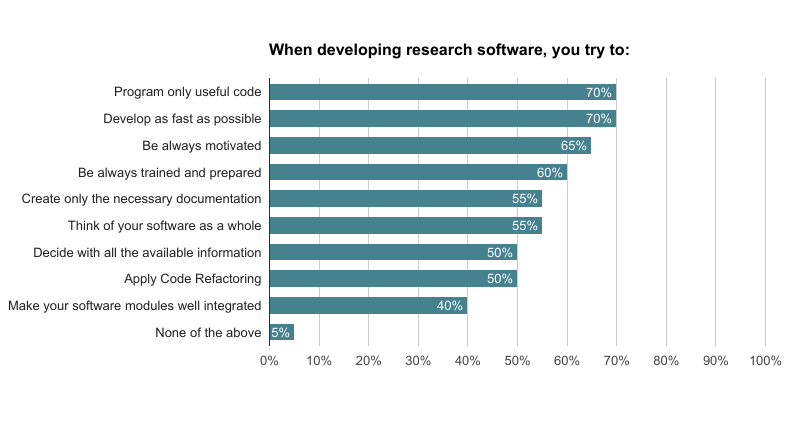}
\caption{Number of participants that agree with each of the LSD principles.}
\label{fig:5}
\end{figure}

Results related to the agreement with the LSD principles are presented in Figure \ref{fig:5}. As the Agile Alliance principles, there is consensus of the participants regarding the LSD principles (8 of the 9 principles have the support of at least 50\% of the participants). However, the correct integration of the software modules does not seem to be so relevant for the researchers, since only 40\% of the participants agree with it. Another important issue to point out is that 5\% of the participants did not agree with any of the LSD principles.

% Regarding the programming languages used by the participants, C/C++ is the most used one, followed by Java, as shown in Figure \ref{fig:6}, whilst Python and M (Matlab) are used by less than 50\% of the participants. Other programming languages that are used by the researchers are Groovy, Assembly, R, JavaScript, PHP and Ruby.

% \begin{figure}[!hbt]
% \centering
% \includegraphics[scale=0.4]{fig6}
% \caption{Programming languages used by the researchers.}
% \label{fig:6}
% \end{figure}

\begin{figure}[!hbt]
\centering
\includegraphics[scale=0.4]{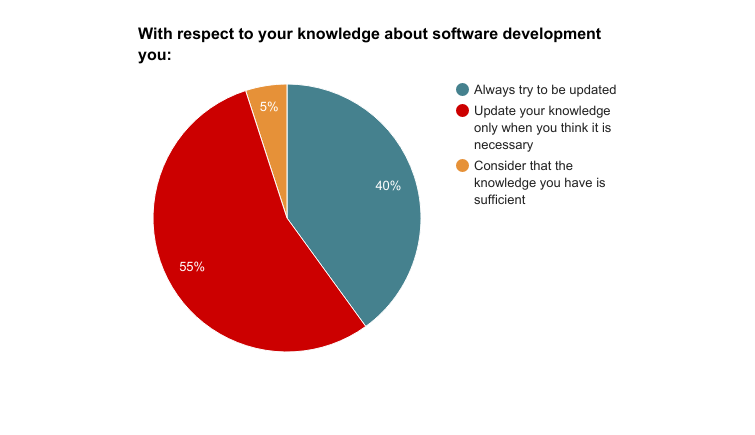}
\caption{Openness to update the knowledge regarding software development.}
\label{fig:7}
\end{figure}

As for the interest for updating their knowledge about software development, which results are shown in Figure \ref{fig:7}, most of the participants (55\%) confirmed that they try to update their knowledge only when it is necessary for a project. A high percentage (40\%) of participants also stated that they always try to keep updated whilst only 5\% considers that it is not necessary. This shows that most researchers recognize that this is a crucial task for doing research.

\begin{figure}[!htb]
\centering
\includegraphics[scale=0.4]{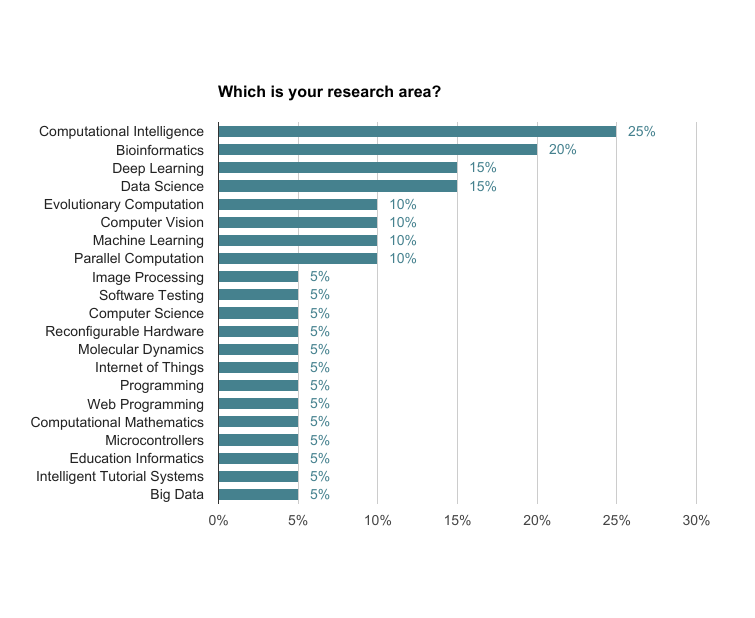}
\caption{Lines of research of the participants. The inferior axis presents the number of researchers working at each field.}
\label{fig:8}
\end{figure}

Although all researchers develop software to a greater or lesser extent for their research projects or experiments, they work in different areas. Figure \ref{fig:8} presents their lines of research. In the survey, this question allowed the researcher to specify his line of research, which is why the figure presents a high number of research areas. The most common line of research is Computational Intelligence: 25\% of the researchers work in that field.

\subsection{Threats to validity}
The number of participants is a threat to validity of this work. We also consider that the backgrounds and research areas of the participants may bias the results of this work, since most were computer scientists. Future works may aim at collecting data from a bigger and more heterogeneous group of participants, they can also include non-Brazilian researchers, in order to obtain more general results.

%\vspace{-12.5pt}
%%%%%%%%%%%%%%%%%%%%%%%%%%%%%%%%%%%%%%%%%%%%%%%% Conclusion %%%%%%%%%%%%%%%%%%%%%%%%%%%%%%%%%%%%%%%%%%%%%%%%
%\newpage
\section{Conclusion}
\label{seq:conclusion}

% What had been done
This work aimed at studying the perception of researchers about the application of agile software development methods for research projects. Since, as far as we are aware of, there are no research has been published regarding the application of agile principles and practices in the research field, this study addressed this issue by constructing and applying a questionnaire to $20$ researchers of diverse backgrounds that need to develop software for their projects.

% Discussion 
%Results obtained in this work show that all of the participants have at least some superficial knowledge about agile methods. Researchers are generally open to apply agile methods since most of them at least intend to apply them in their projects. As for agile practices, the participants usually apply at least their core concepts when developing software. There is also a general agreement regarding Agile Principles (Agile Alliance and LSD) and regarding to updating the software development knowledge.

Results show that researchers are open to apply Agile Software Development methods for their projects and that they already apply at least the core concepts of some agile practices. They also agree with the agile principles and exhibit openness to updating their knowledge regarding software development. The leading agile method is Scrum, followed by eXtreme Programming. The perception of researchers on applying agile software development methods on their projects is positive. There is a consensus regarding the benefits of agile development, since the participants demonstrated openness to apply agile practices and principles.

Taking these results into account, we can conclude that, in general, researchers consider Agile Methods as a viable option to develop software for their projects, since the software they develop usually have dynamic requirements and most of them already apply agile practices even without having formal knowledge about them. This positive response of the researchers may allow to carry more studies aiming at evaluating the application of ASD methods in research projects and their impact on the development process when compared to traditional methods.

% Future work
For future works, the number of participants can be augmented along with the questions contained in the questionnaire. Another survey structure can also be applied by building one questionnaire for each agile method in order to obtain more specific results for each. The heterogeneity of the participants is another objective that can be pursued in future works, since the results presented in this document correspond to those obtained Brazilian researchers only. Thus, the survey can be also applied to participants from another countries.

% \section*{Acknowledgements}
% N.Aquino thanks the Organization of the American States, the Coimbra Group of Brazilian Universities and the Pan American Health Organization.

%\section*{References}

\bibliographystyle{unsrtnat}
\bibliography{main}

\end{document}